\documentclass[12pt]{article}
\usepackage{latexsym}
\usepackage{psfig}
\usepackage{epsfig}
\usepackage{amsmath}
\usepackage{citesort}
\usepackage{amssymb}

\begin{document}

\def\cH{{\cal H}}

\begin{center}
{\large \bf Relaxation times in the ASEP model using a DMRG method}

\vskip 1cm
{ Zolt{\'{a}}n Nagy$^a$, C{\'{e}}cile Appert$^a$, and Ludger Santen$^{a,b}$}

\vskip 0.1cm
\small{${}^a$ Laboratoire de Physique Statistique
\footnote{Laboratoire associ\'e aux universit\'es Paris 6,
Paris 7 et au CNRS}\\
Ecole Normale Sup\'erieure, \\
24 rue Lhomond, F-75231 PARIS Cedex 05, France\\
${}^b$ Theoretische Physik\\
Universit\"at des Saarlandes\\
66041 Saarbr\"ucken, Germany\\
appert@lps.ens.fr, zoltan.nagy@polytechnique.org, santen@lusi.uni-sb.de}

\date{\today}

\end{center}

\begin{abstract}
We  compute the largest relaxation times for the totally asymmetric 
exclusion process (TASEP) with open boundary conditions
with a DMRG method.
This allows us to reach much larger system sizes than in
previous numerical studies. We are then able to show that 
the phenomenological theory of the domain wall indeed 
predicts correctly the largest relaxation time for large systems.
Besides, we can obtain results even when the domain wall
approach breaks down, and show that the KPZ dynamical exponent
$z=3/2$ is recovered in the whole maximal current phase.
\end{abstract}

\vskip 1cm
{\em PACS numbers: 05.40.-a, 05.60.-k, 02.50.Ga}

\vskip 1cm
{\em Key words: asymmetric exclusion process, density-matrix renormalization, 
dynamical exponents}

\vskip 1cm

\section{Introduction}

Several many particle systems are characterized by a steady mass transport.
Examples for this kind of systems can be found in
biological transport \cite{MacD69} or vehicular traffic \cite{review}.
 From a theoretical point of view these processes are of particular 
interest, because they exhibit generic non-equilibrium behavior. 
Due to the large number of important applications, many microscopic 
models for particle transport have been suggested in recent years 
\cite{review,privman,marro}.

Among  these, the most important microscopic model for non-equilibrium 
particle transport is the so called asymmetric exclusion 
process (ASEP) \cite{gunter,spohn}. In this model, particles jump on a
one-dimensional lattice, either to the right (with probability $p\; dt$)
or to the left (with probability $q\; dt$), if the corresponding
sites are empty.
The model shows a number of generic effects \cite{Krug91} that 
are characteristic for non-equilibrium particle transport and 
maintain for the more specialized variants of the model 
\cite{Popkov99,Popkov00}. 
At the same time the ASEP is simple enough to obtain several exact results
for the system, which is of great importance, because the general 
theoretical framework of non-equilibrium physics is less developed.
Exact results exist, e.g. for the stationary state of the system 
with periodic \cite{spohn} 
and open boundary conditions \cite{Derr92,DEHP,Gunter93}.
The case of open boundary
conditions is of special interest because 
one observes boundary induced phase transitions \cite{Krug91}.
In this paper, we restrict ourselves to the totally asymmetric exclusion
process (TASEP), that is $q=0$. This case includes the most important
phenomena but simplifies considerably the discussion of the model.
Open 
boundary conditions are implemented by two particle reservoirs 
that are coupled to the chain. The capacities of the reservoirs determine,
together with the bulk hopping rates, the actual state of the system.

The complexity of the open system is also reflected in the 
mathematical structure of the stationary solution.
While the steady state of the periodic system is given by a simple 
product measure \cite{spohn}, it is highly non-trivial for the open system. 
Nevertheless, it can be calculated and it is possible to obtain
several non-trivial quantities, e.g. current- or density fluctuations in the 
stationary state \cite{Derr92,DEHP,Gunter93}, or  large deviation
functions \cite{Derrida}.

The dynamic properties of the open system are, however, more puzzling. 
Exact analytical results for the largest relaxation time $\tau$
and the corresponding 
dynamical exponent $z$ are so far only possible for the periodic chain 
by applying the Bethe ansatz \cite{Gwa,Kim}. For the open chain 
estimates for $\tau$ can be obtained from a phenomenological approach, 
that models directly the dynamics of the boundary layer separating 
the high and low density domains imposed by the particle reservoirs
\cite{Kolo98}.
It has been shown, that this approach gives asymptotically correct results
in a certain parameter regime \cite{Belitzky}.
 For finite systems, as well as for 
general in- and output rates, relaxation times have to be calculated 
numerically.

This has been done by U. Bilstein and B. Wehefritz \cite{Bilstein}, and 
later by M. Dudzi\'nsky and G.M. Sch\"utz \cite{Dud} who calculated the 
relaxation times by using exact diagonalization. This method 
is, however, restricted to very short chains (less than twenty),
that are not in the asymptotic limit. In this work we used 
the density matrix renormalization group (DMRG) technique
\cite{white}, that 
enables us to calculate the relaxation times for  much larger system 
sizes, compared to \cite{Bilstein,Dud}. By treating large 
chains we have obtained  more conclusive results for the 
dynamical exponent $z$ in the maximal current phase and 
could also achieve convergence between the exact numerical 
values of the relaxation times and the estimates of the phenomenological 
approach.

The paper is organized as follows. In the next section we discuss 
briefly the relevant physical concepts and the applied numerical 
techniques. In the third section we show the comparison of the domain
wall predictions for finite systems with our results. Section 4
is devoted to the special case of the disorder line $\alpha+\beta=1$.
This section is followed 
by a discussion of the dynamical behavior when approaching the 
phase boundaries as well as in the maximum current phase.
 
\section{Analytic predictions and the DMRG method for non-equilibrium systems} 

\subsection{The TASEP with open boundary conditions}

For self-containedness we will repeat the definition of 
the model. The TASEP is 
defined on a one-dimensional lattice with $L$ sites. 
The boundary sites of the chain are coupled to two particle
reservoirs, one reservoir on the left that controls the particle input 
and a second on the right that governs the output of particles. 

We regard the process in continuous time (see~\cite{raj} for  
a comparison of the different update procedures), which 
corresponds to a random sequential update in computer 
simulations. If a link between sites $i$ and $i+1$ is 
selected, a particle located at $i$ moves to site $i+1$
if site   $i+1$ is empty (for convenience we set the hopping rate to
one). In case of choosing the link $(0,1)$ one introduces 
a particle with probability $\alpha$ if the first site is 
empty. Finally a particle may leave the system with 
probability $\beta$, if the link $(L,L+1)$ is chosen.     

Each configuration $\sigma $  can be written in
terms of boolean lattice gas variables $\sigma_i$, i.e. $\sigma_i = 0 (1)$ if
the site is empty (occupied). If we introduce an orthonormal basis
$|\sigma \rangle=|\sigma_1,...,\sigma_L \rangle$ in the $2^L$-dimensional
configuration space, we can define the probability
vector  $|P(t)\rangle$ as
$|P(t)\rangle=\sum_{\{\sigma\}} P(\{\sigma\},t) |\sigma\rangle$.
The time evolution of $|P(t)\rangle$ is determined by means
of the master equation,  that can be written
as a  Schr\"odinger equation in imaginary time \cite{gunter,raj}:
\begin{equation}
\label{schroed}
\frac{\partial}{\partial t} |P(t)\rangle =-\cH \, |P(t)\rangle ,
\end{equation}
where $\cH$ denotes the stochastic Hamiltonian. The matrix
elements of $\cH$ are the rates $w(\sigma  \rightarrow \sigma ')$ for a
transition $\sigma \to \sigma '$.
Explicitly $\cH$ is given by
$\langle \sigma|\cH|\sigma'\rangle=-w(\sigma ' \rightarrow \sigma)$
for the off-diagonal elements ($\sigma\neq\sigma'$) and by\\
\noindent
$\langle \sigma|\cH|\sigma\rangle=\sum_{\{ \sigma \} \ne \{ \sigma'
  \}}w(\sigma \rightarrow \sigma')$
for the diagonal elements.

\begin{figure}[h]
\centerline{\epsfig{figure=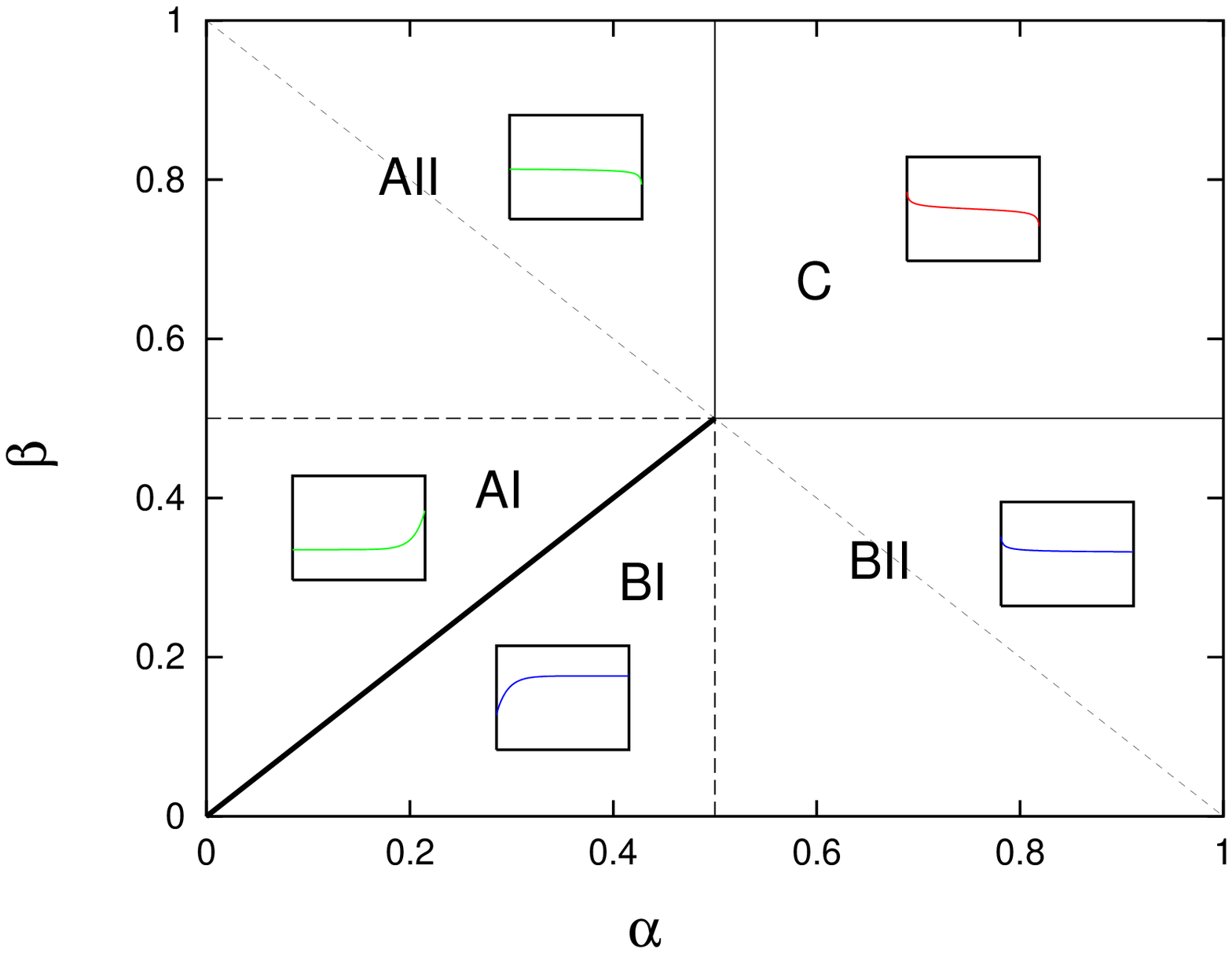,height=9cm}}
\caption{Phase diagram of TASEP with random
sequential update, depending on the input rate $\alpha$ and
output rate $\beta$. The insets show typical density profiles.
Phase transitions are indicated by the solid lines.
Along the diagonal dashed line $\alpha + \beta = 1$, corresponding
to the disorder line of the model, the density profile
is flat.
Finally, the transitions between the subphases AI / AII
(respectively BI / BII) are marked by dashed lines.
}
\label{asep_phase}
\end{figure}

The exact stationary solution of the master equation 
can, for general $\alpha, \beta$, be written as a product of
infinite dimensional matrices \cite{DEHP}. This solution allowed to 
calculate the $\alpha,\beta$ dependence of the stationary 
quantities, e.g. the average flux. The results are summarized 
in the phase diagram of the system shown in fig.~\ref{asep_phase}. 
Three phases can be distinguished by means of a different functional 
behavior of the flow. In the low density phases AI and AII, i.e. 
for $\alpha < \beta$, $\alpha <1/2$,  the flow $J= J (\alpha)$ is given by 
$J = \alpha(1-\alpha)$ and analogously in the high density phases BI and BII
by  $J = \beta(1-\beta)$. If both  $\alpha > 1/2$ and  $\beta >  1/2$
(phase C)
the capacity of the particle reservoirs exceeds the capacity of the
chain. Then the flux is independent of $\alpha$ and  $\beta $ and
given by $J = 1/4$. This phase is called maximal current phase.

Both the high and low density phases are
divided into two subphases.
In phase AI and BI, the capacities of both reservoirs are
below the chain capacity. In phase AII (BII), only the capacity
of the exit (entrance) exceeds the chain capacity.
This has, e.g. consequences for the asymptotics of the
density profile \cite{Gunter93}, and, a question
we address in this article, possibly also for the dynamics
of the chain. Another important line is given for $\alpha + \beta=1$.
On this one-dimensional line the stationary solution is much simplified,
i.e. it is given as product measure. 

Now we shall summarize known results for the relaxation times.
For the periodic system, exact results
for the energy-gap,  i.e. the inverse of the largest relaxation 
time $\tau_1$,  have been obtained by Bethe-ansatz techniques
\cite{Gwa,Kim}. The largest relaxation time scales for any density 
of the system asymptotically as $\tau_1 \sim  L^z$, where 
$z= 3/2$.

As translational invariance is lost in the open system,
the technique does not apply.
Besides, the dynamics is profoundly modified by the presence
of the two particle reservoirs, which impose the coexistence
of two domains into the system.
In case of the phases $AI,BI$, it is known that the two
domains imposed by the reservoirs have a simple factorized structure.
In this parameter regime the so-called domain
wall (DW) theory can be applied \cite{Kolo98}.  The
DW theory uses a coarse grained description of the dynamics of the
process: Each particle reservoir that is coupled to the chain
imposes independently a domain of a given constant density
$\rho_L$ and $\rho_R$. The two domains are separated by a localized domain
wall, which performs a biased random walk. The bias is
due to the different capacity of the two reservoirs and can
be calculated simply by using the conservation of mass.
In a finite system the position of the domain wall is confined
between two reflecting walls. The description of the process
allows to calculate the stationary as well as the fully time
dependent probability distribution of the domain wall positions \cite{Dud}.
Then, it is in particular possible to estimate the largest relaxation times
of the system, that are given by
\begin{equation}
\tau_n = \left[ D^+ + D^- -2\sqrt{D^+  D^-} \cos(\pi n /(L+1))\right]^{-1}
 \qquad n = 1,2,\dots
\label{eqdw1}
\end{equation}
with
\begin{equation}
D^+  = {\beta (1-\beta) \over 1 - \alpha - \beta} ; \;\;\;
D^-  = {\alpha (1-\alpha) \over 1 - \alpha - \beta}.
\label{eqdw2}
\end{equation}

These results are valid for $\alpha, \beta < 0.5$. 
The remaining parameter space has to be explored numerically. 

\subsection{The DMRG method for stochastic models}

In the previous section we have shown that the 
master equation can be rewritten as a Schr\"odinger 
equation in imaginary time.
Now, in order to calculate the longest relaxation time
of the system, one has to calculate the two lowest eigenvalues 
of eq.~(\ref{schroed}) (the first eigenvalue is trivially zero,
with an eigenvector corresponding to the stationary state).
The eigenvalue calculations have 
to be done with very efficient diagonalization methods, in order to reach 
sufficiently large system sizes. This is possible by applying
the DMRG method \cite{whitereview}.

This method was first developed to study
the properties of strongly correlated electrons
\cite{white}. It has recently been generalized in order to 
treat stochastic many particle systems \cite{carlon1,carlon2}.

The idea of the DMRG is the following. One starts with a small system (here 12 sites)
that can be diagonalized with standard numerical methods.
Then one performs a large number of renormalization cycles in order to increase
the system size.
At each renormalization cycle, first the system is enlarged by adding
2 sites in the bulk of the system, i.e. in a place where it is less likely that the
eigenmodes will be perturbed. 
The largest eigenvalues of the corresponding enlarged Hamiltonian are
computed.
These eigenvectors are used in order to construct the density matrix
of the system, which will be used in the next stage.
Second, in order to avoid an 
exponential growth of the hamiltonian,
 a projection onto the ``most
 important'' modes has to be done.
It has been shown that the best choice is to keep the $m$ leading
eigenvectors of the density matrix~\cite{white}. 
The parameter $m$ has to be chosen small enough in order
to allow a fast calculation of the eigenvectors, but
large enough in order to obtain a high numerical precision.
The error due to each truncation is well controlled \cite{whitereview}.

Although no fundamental differences between the DMRG method for 
stochastic and quantum mechanical systems exist, one has to 
make a certain effort in order to overcome the numerical 
difficulties. This has been done following the suggestions 
of ref.~\cite{carlon1,carlon2}. We now discuss some details
of our implementation.

The main difference between 
the quantum mechanical and stochastic problem is that the 
stochastic hamiltonian is not hermitian. Therefore, the 
calculation of the eigenvectors of the hamiltonian
required at each renormalization
step cannot be done by 
standard diagonalization techniques (as the basic L\'anczos
or Davidsson algorithms) but one has to apply, e.g. 
the Arnoldi method or L\'anczos for non-symmetric matrices.
These methods are quite efficient but, 
compared to their analogues for symmetric matrices, less 
stable~\cite{golub}. We use the Arnoldi method, that turned 
out to be numerically more stable than the L\'anczos method for 
non-symmetric matrices. Nevertheless the method is numerically 
well controlled, because the accuracy of the calculated 
eigenvector can be obtained via the residual norm, i.e.
we check that it is indeed an eigenvector.
Actually, we have used the Arnoldi method in an iterative way,
where the initial guess is the result of the previous Arnoldi run,
until the desired precision is reached.

Apart from being non-hermitian, there is another difference 
between stochastic and quantum mechanical problems. It is 
well known that the lowest eigenvalue of stochastic 
hamiltonians is always zero, corresponding to the existence
of a stationary state,
and that the corresponding 
left eigenvector $|0_l\rangle$ has all its coordinates equal to 1.
Therefore the first task will be
to calculate the right eigenvector $|0_r\rangle$. As we are interested in 
the asymptotic dynamics of the system, we have to calculate 
the energy gap G(L) as well, i.e. the next smallest eigenvalue
and the corresponding left and right eigenvectors $|1_l\rangle$ and $|1_r\rangle$.
Then the longest relaxation time $\tau = \tau_1$ will be given by
\begin{equation}
\tau = G(L)^{-1}.
\end{equation}

After the calculation of these eigenmodes, one has to
eliminate the least important degrees of freedom of
the system, in order to keep the
system at a manageable size.
It has been shown that the relevant degrees of freedom
are the eigenmodes associated to the largest eigenvalues
of the density matrix $\hat{\rho}$.
The latter is defined depending on how many eigenmodes we
need to calculate. Here, for the calculation of the gap,
we take
\begin{equation}
\hat{\rho} = {1\over 2} \hat{Tr}\left\{ |0_l\rangle\langle0_l| + |0_r\rangle\langle0_r| + |1_l\rangle\langle1_l| + |1_r\rangle\langle1_r| \right\}.
\end{equation}
Note that it is always possible to construct a symmetric 
form for the density matrix. Therefore, standard algorithms can be applied
 in order to calculate all eigenvalues and eigenvectors of $\hat{\rho}$ 
as for quantum systems.

Another trick that allowed to gain accuracy on the gap calculation
is to compute first the right eigenmode $|0_r\rangle$ for the 
fundamental state (the left is known), and then to define a 
new hamiltonian
\begin{equation}
H^\prime(\Delta) = H + \Delta |0_r\rangle\langle0_l|.
\end{equation}
This hamiltonian $H^\prime$ has, apart from the groundstate, 
the same spectrum as the original hamiltonian $H$.
The gap is now given by the fundamental state of $H^\prime$ \cite{carlon2}.

Note that one requirement of the DMRG method is that the requested
eigenvalue is well separated from the next one.
This may not be true when the system size increases, and then
the calculation becomes instable.
The system size limitations come from these instabilities
rather than from runtime or memory requirements.
So, the DMRG method either gives very accurate results for the eigenvalues,
or does not converge at all.

Finally we want to mention another particularity of the system.
As rules are different
at each end of the system (input or output),
we cannot use the left/right symmetry of the system
as people do for closed systems, and the left and right
parts of the system have to be computed alternatively.

\section{Test of the Domain Wall Theory at Small System Sizes}

 The comparison between DW predictions and the exact analytical 
for the stationary state shows that they agree in the limit of 
large system sizes and $\alpha, \beta < 1/2$ \cite{Belitzky}. 
For finite systems, however, the coarse grained picture slightly 
deviates from the exact results. 

In case of the dynamical properties, so far no exact analytical 
results exist. Therefore the DW predictions have to be checked 
with numerical methods.

In a previous work, we have compared time-dependent density profiles
in a non-stationary regime, and found good agreement
with the domain wall predictions \cite{santen_appert}.
Here, we calculate explicitly the relaxation times of the process, in
order to compare with the predictions of the DW method.
The same has been done already by using the non-symmetric Arnoldi method
\cite{Bilstein,Dud}, but the 
 system sizes that  one can treat by the  non-symmetric Arnoldi
method are too small ($L \le 16$ in \cite{Dud}) in order to obtain
convergence with the DW predictions.
Our aim is to improve this convergence thanks to the ability of
DMRG to treat larger systems.

\begin{figure}[h]
\centerline{\epsfig{figure=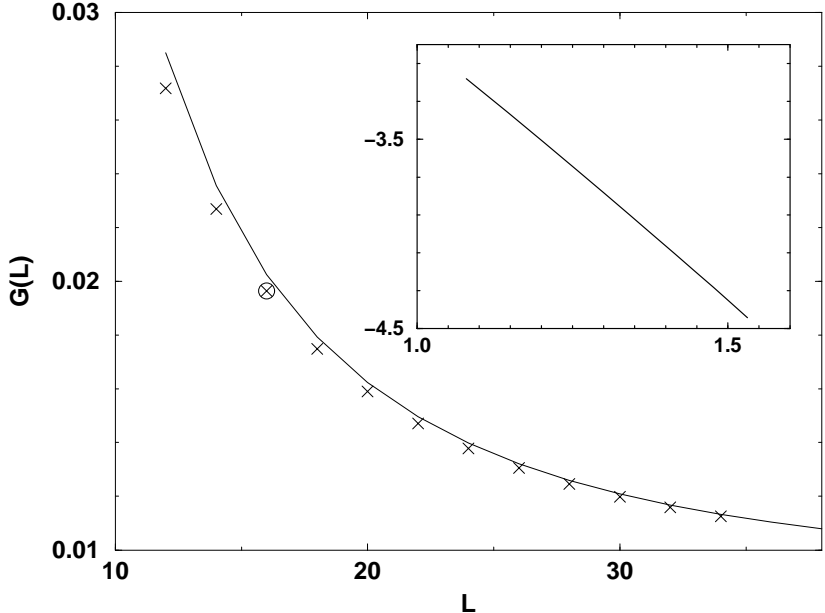,height=9cm}}
\caption{Gap as a function of the system size,
for $\alpha = 0.2$ and $\beta = 0.1$.
The solid line indicates the domain wall prediction,
while symbols represent the DMRG results.
The circle was obtained in \protect{\cite{Dud}} for L=16.
Error bars for the DMRG results are not given, as they
are much smaller than the size of the symbols.
The inset shows in a log-log plot the difference between
the two curves.
}
\label{fig1}
\end{figure}

First we consider cases with both $\alpha$ and $\beta < 1/2$, where 
the domain wall theory is expected to work well, though this was not
fully verified by the exact numerical calculation, due to size 
limitations. Fig.~\ref{fig1} shows the comparison of the inverse 
relaxation time with the domain wall predictions in the high 
density phase ($\alpha = 0.2$, $\beta = 0.1$). For comparison 
we also included the result for $L=16$ obtained in \cite{Dud}, which
perfectly coincides with the DMRG calculations. Obviously
the theoretical predictions and the numerical results are in good
agreement, if the length of the chain reaches thirty sites.

\begin{figure}[h]
\centerline{\epsfig{figure=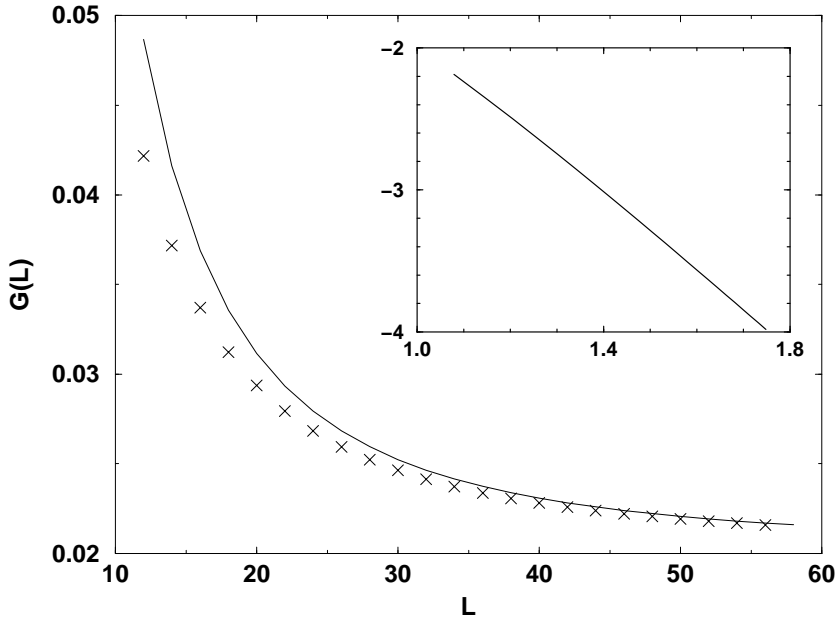,height=9cm}}
\caption{Gap as a function of the system size,
for $\alpha = 0.4$ and $\beta = 0.2$.
The solid line indicates the domain wall prediction,
while symbols represent the DMRG results.
The insets shows in a log-log plot the difference between
the two curves.
}
\label{fig2}
\end{figure}

For $\alpha = 0.4$ and $\beta = 0.2$,  the DW predictions 
for small system sizes are slightly less accurate,
as shown in figure
\ref{fig2}. This could be expected, as the density difference between
the two domains $\Delta \rho = 1-\beta-\alpha$ is smaller and
therefore the width of the shock larger than
in the previous case. If, however, the shock touches 
the boundaries the approximation of microscopic 
states by $\Theta$-functions is less appropriate.

For $\alpha$ or $\beta$ larger than 0.5, the coexistence is not between 
a high- and low-density domain, but, e.g. for $\alpha>1/2$, between a
maximal current and high-density domain. This case is much more complicated 
than the coexistence of domains in the phase $AI$ or $BI$. It is known
from the exact results, that long ranged correlations exist 
in the maximal current regime, which implies that the maximal 
current domain has no simple structure, and cannot be easily described.
 
A naive attempt to describe the maximal current domain,
which uses the fact that the flux is a constant
$J=1/4$ and would lead to exponentially decaying
density profiles, is to build the density profile iteratively from the
boundary with a mean field formula like $j=\rho_i (1 - \rho_{i+1})$.
But this yields a wrong density profile.
It was also suggested \cite{Dud} to use the exact results of the
density profile.
However, this approach has intrinsic shortcomings.
First the DW is no longer self-contained and, second,
it is no longer possible to specify the relaxation times in a
closed form, because the hopping rates of the random walk are now
site-dependent.
But above all, the description of the system states as the
juxtaposition of two phases with an independent boundary layer
between them breaks down.
Indeed, the wall dynamics is coupled to the
whole structure of the maximal current phase, due to the long
ranged correlations.
 
So we explore this parameter regime guided by two
questions: $(i)$ Do we obtain a finite relaxation time in the
phases $AII /BII$ in the limit of large system sizes? $(ii)$
And, if it is the case, is it possible to model the maximum
current phase as a flat domain with an effective density $\rho_{eff}$?

\begin{figure}[h]
\centerline{\epsfig{figure=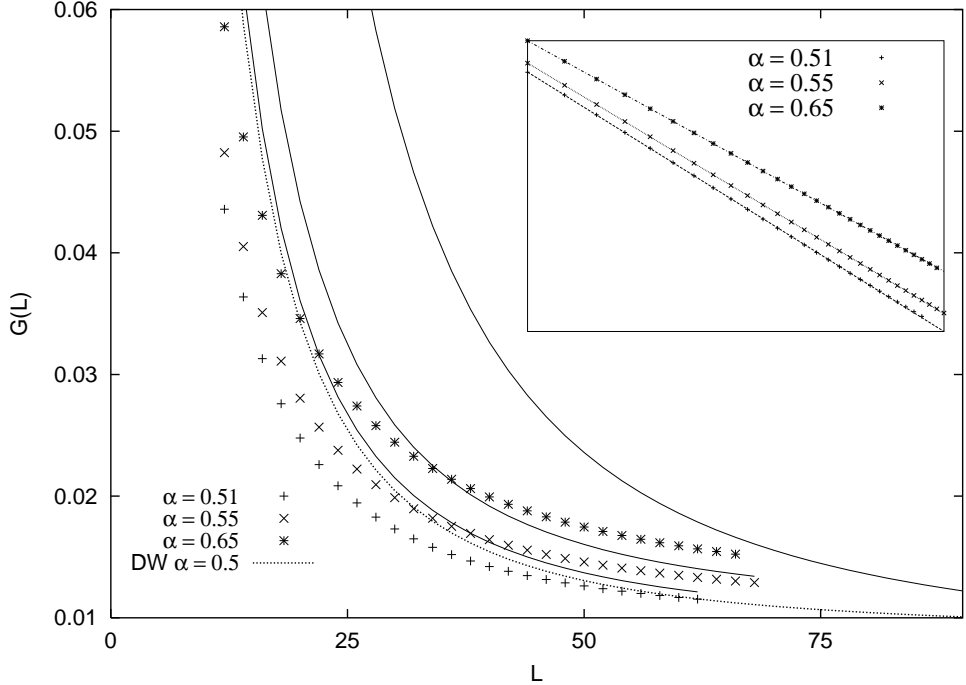,height=9cm}}
\caption{Gap as a function of the system size,
for $\alpha = 0.51$, $0.55$, and $0.65$, and $\beta = 0.3$.
The solid lines indicate the domain wall prediction,
while the symbols represent the DMRG results.
The domain wall prediction for $\alpha = 0.65$ is the upper curve.
The dotted line gives the domain wall prediction
calculated for $\alpha=1/2$.
The DMRG results $y_\alpha(L)$ were interpolated using an algebraic fit function
$f(L) = a L^b + c$. The inset shows $\log[f(L) - c]$ as a function
of $\log L$ (symbols) in comparison with $\log[y_\alpha(L) - c]$
(solid lines).
}
\label{fig_abig}
\end{figure}

First we checked the domain wall predictions close 
to the transition line at $\alpha = 0.5$, i.e. 
for  $\alpha = 0.51$ and $0.55$ (see figure \ref{fig_abig}).
In this parameter regime one already has long ranged correlations 
in the maximal current domain, but at the same time the magnitude 
of the deviations from a flat domain
is rather small. Our results indicate, that within 
this parameter regime the DW formulas (\ref{eqdw1}-\ref{eqdw2}) still
lead to satisfying results for quite large system sizes.
On the other hand, the DW prediction with $\alpha=1/2$,
which was conjectured at first to be the proper one for
any $\alpha>1/2$ \cite{Dud}, clearly underestimates
the gap for any system size.

For larger values of $\alpha$ (e.g. $\alpha=0.65$ in
figure \ref{fig_abig}), 
the algebraic corrections of the density profile become 
non-negligible.
Equations (\ref{eqdw1}-\ref{eqdw2}) do not give anymore
a good estimate.

 Next, we checked whether our results are compatible with
a finite relaxation time in the limit $L \to \infty$. 
Therefore we interpolated our results for $G(L)$ using 
the functional form $f(L) = a L^b + c$. We find a very 
good agreement between the fits and our numerical data,
with a finite value of $c$ (see the inset of fig.~\ref{fig_abig}), 
which indicates that the relaxation times remain finite. Our 
estimates for the relaxation times are  $\tau(\infty) = 116.(6)$
for $\alpha = 0.51$,
$\tau(\infty) = 101.(7)$ for $\alpha = 0.55$,
and $\tau(\infty) = 96.(1)$ for $\alpha = 0.65$.

A finite relaxation time for $L\to \infty$ 
implies at the same time, that the dynamical exponent
vanishes. This can be checked systematically by 
calculating the size dependent dynamical exponent given as:
\begin{equation}
z(L) = -\frac{\ln\left[ G(L+2) \right] - \ln\left[ G(L) \right]}{\ln(
  L+2 ) - \ln( L )}.
\label{eq:z_fs}
\end{equation}

\begin{figure}[h]
\centerline{\epsfig{figure=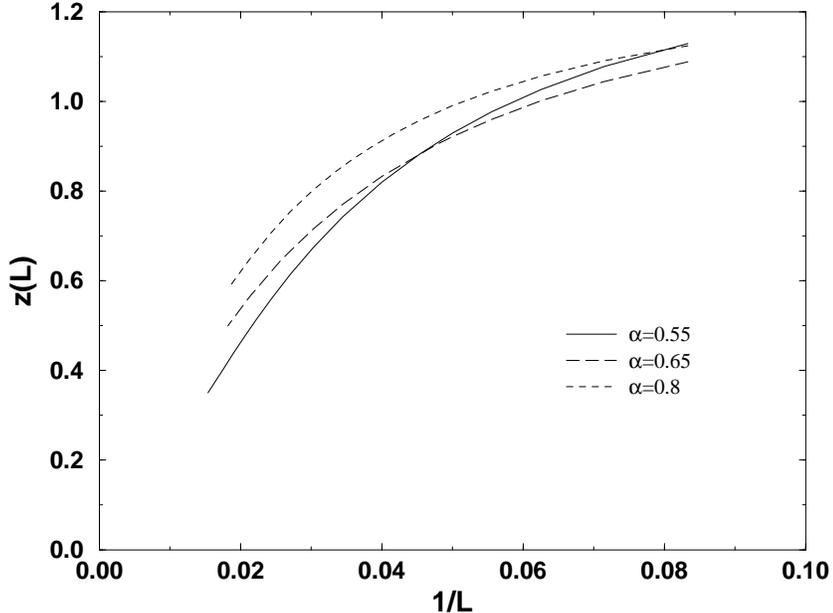,height=9cm}}
\caption{Size dependent dynamical exponents for different 
values of $\alpha$ ($\beta = 0.3$). The results are 
compatible with $z(\infty) = 0$.}
\label{fig_expaii}
\end{figure}

Figure \ref{fig_expaii} shows that our  results are compatible with  $z=0$ for 
all values of $\alpha$  we took into account. 
So these  results for $\alpha> 0.5$  further support that 
the relaxation times are finite within these subphases.
 
Now, one can simply take the values $\tau(\infty)$ in order to 
calculate an effective density of the maximal current domain.  
This procedure leads to DW predictions for small systems, which 
do not agree very well with our numerical findings. We thus 
believe that the non-trivial structure of the 
maximal current domain must be considered in modeling the
domain wall motion.

\section{Disorder line}

For general $\alpha,\beta$ it has been shown that the stationary 
weights are given as products of infinite dimensional matrices.
Nevertheless a one-dimensional line in the parameter space
exist, i.e. $\alpha + \beta = 1$, 
where the stationary weights are  simple products
as for the periodic systems.
These product states are in analogy with disordered states
of  quantum spin models. At the disordered line the system is 
homogeneous, i.e. one can not identify two different domains. 
This  implies that the relaxation at this particular 
line is not governed by the motion of the domain wall.

Then it is an open question to know whether dynamical properties
are
changed on this line  and how the dynamical properties compare to 
the periodic system.
For the periodic system, Bethe ansatz predicts the dynamical
exponent $z = 3/2$ \cite{Gwa,Kim}. Besides, the first 
non-zero eigenvalue has an imaginary part 
except for the density $1/2$. 
The divergent relaxation times of the periodic system are,
 however, related to its
translational invariance \cite{beijeren}. 

By contrast, for the open
system it is expected that the relaxation times are finite
if $\alpha \neq 0.5$, because the density fluctuations spread
with a nonzero drift velocity \cite{Krug91}.

First, we have applied the DMRG method to the special point
$\alpha = \beta = 0.5$, where the drift velocity of an 
excess density vanishes.

\begin{figure}[h]
\centerline{\epsfig{figure=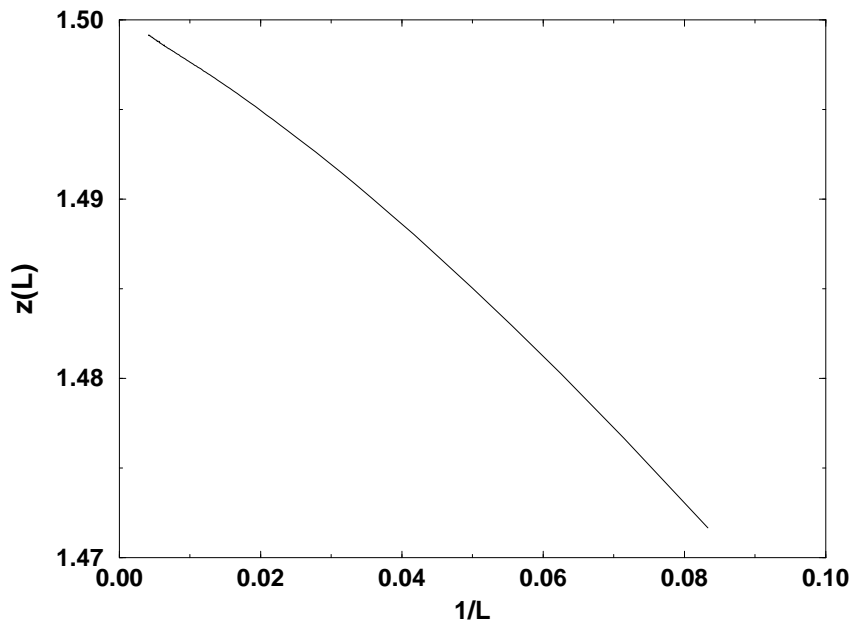,height=9cm}}
\caption{Dynamical exponent $z$ as a function of the system size,
for $\alpha = \beta = 0.5$.}
\label{fig_kpz}
\end{figure}

Fig. \ref{fig_kpz} shows that our numerical results for $z(L)$, obtained up
to $L=244$, agree with
the exponent $z=3/2$ predicted for the periodic system.

\begin{figure}[h]
\centerline{\epsfig{figure=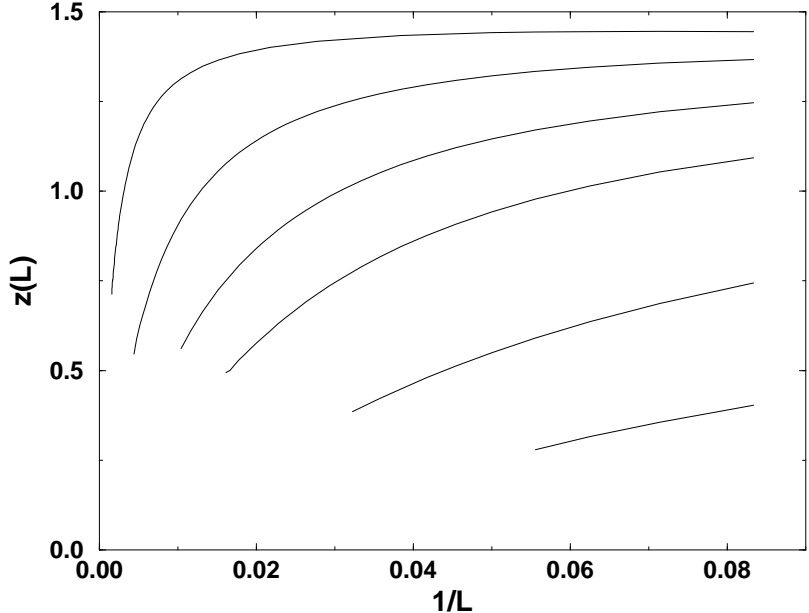,height=9cm}}
\caption{Dynamical exponent $z$ as a function of the system size,
along the disorder line. The curves from bottom to top correspond
to $\alpha = 0.1$, $0.2$, $0.3$, $0.35$, $0.4$, and $0.45$.
Error bars are of the same order as the line thickness.
}
\label{fig_mf}
\end{figure}

We have also explored the remaining part of the disorder
line. We present our simulation results in fig. \ref{fig_mf}.
They seem to exclude the case $z = 3/2$, and  indicate
that the relaxation time would be finite in the large $L$
limit ($z(\infty)=0$). 

 Although these results are consistent with the 
expected relaxation behavior,  we cannot exclude that a 
branch corresponding to  complex eigenvalues could cross 
our solution when $L$ becomes large. 
The DMRG method does not allow to distinguish whether the
numerical instabilities are due to the intersection or
the convergence of two eigenvalue branches.

Nevertheless, we state that our numerical results are in accordance 
with a finite relaxation time and, therefore, consistent with the physical
picture which was presented in \cite{Krug91}. Further support for 
this scenario comes from its position in the phase diagram.
The disordered line touches a phase boundary only in a single 
point, i.e. for $\alpha = \beta =1/2$. All other points 
of the disordered line are located {\it inside} a phase 
where the relaxation times of the system are finite.
Therefore,
it would be rather counter-intuitive to observe a diverging 
relaxation time for these parameters.

\section{Phase transitions and the maximal current phase}

If the capacity of both particle reservoirs exceeds the capacity
of the chain, the maximal current phase is realized. In the 
maximal current phase both localization lengths are infinite. 
This divergence of length scales leads, e.g. to an algebraic slope of the 
density profiles. 

In this phase, previous numerical calculations by Bilstein et al 
\cite{Bilstein}
for $L \le 20$ extrapolated to large systems seemed to indicate an exponent
$3/2$ in the whole maximal current phase.

\begin{figure}[h]
\centerline{\epsfig{figure=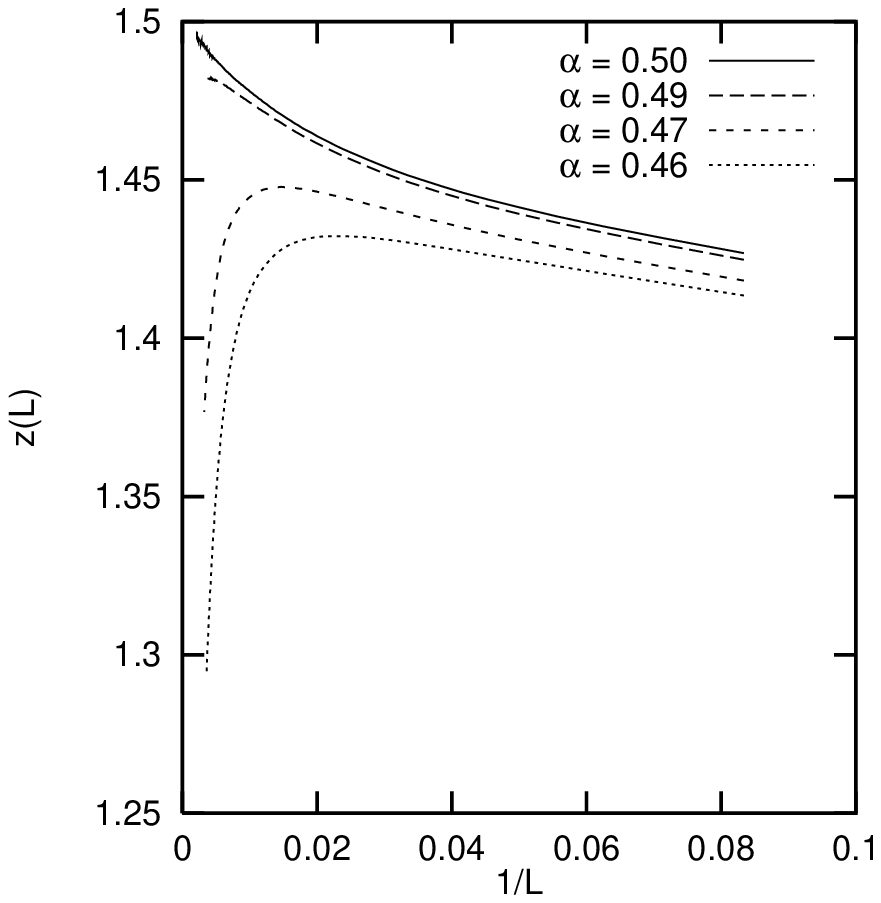,width=0.49\linewidth} \epsfig{figure=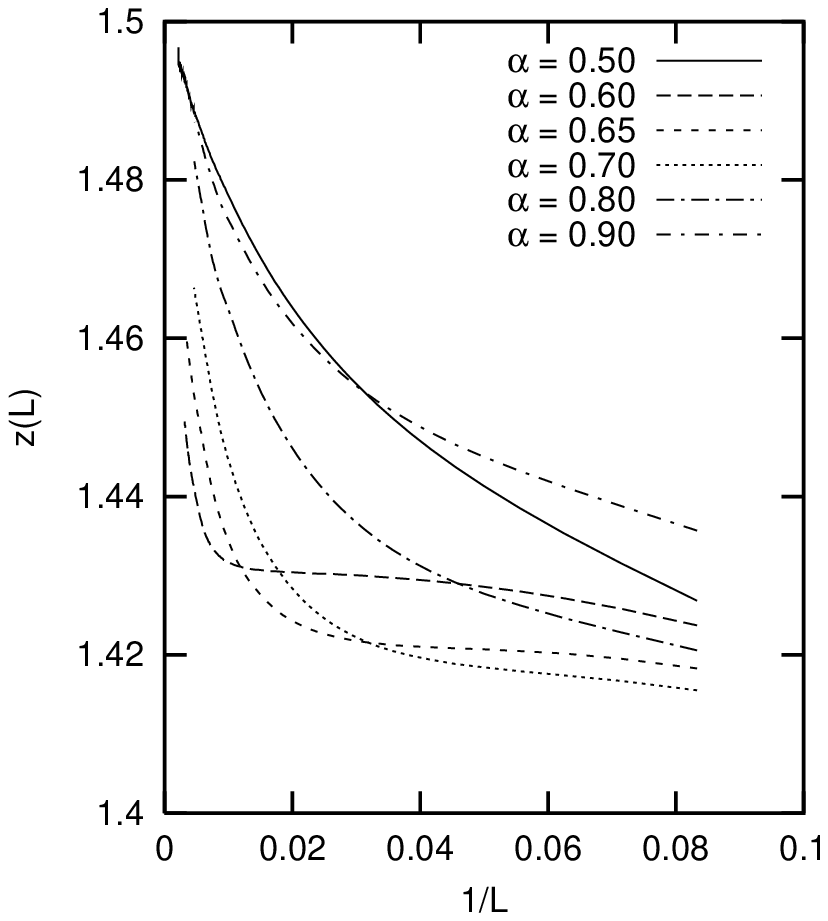,width=0.49\linewidth}}
\caption{Dynamical exponent $z$ as a function of the system size,
for $\beta = 0.8$.
The left figure presents results obtained in the AII phase, where
the maximal current phase coexist with a high density phase.
For $\alpha=0.5$, we are exactly on the 2nd order transition line.
The right figure shows results within
the maximal current phase.
}
\label{fig_mc}
\end{figure}

We have computed the dynamical exponent in various points
in the maximal current phase. Results are presented
in figure \ref{fig_mc}, for a constant $\beta=0.8$,
$\alpha$ being varied through the second order transition ($\alpha=0.5$)
and inside the maximal current phase
(we have checked that of course, results are the same when $\alpha$
and $\beta$ are interchanged).
We clearly see a transition when the maximal current phase
is entered.
On the transition line $\alpha=0.5$, we recover $z=3/2$
in the large system size limit.
Besides, from our numerical results, the infinite size dynamical exponent
seems to be equal to $3/2$ in the whole maximal current phase,
confirming the extrapolation in \cite{Bilstein}.
 
\section{Conclusion}

The use of a DMRG approach to compute the largest relaxation
time in the TASEP model has allowed to give new strong
evidence of the validity of the domain wall picture
when reservoirs control the flow.

For $\alpha$ or $\beta$ larger than 1/2,
the capacity of one of the reservoirs becomes
larger than the chain capacity, and 
one of the coexisting domains is then a maximal
current phase.
Our numerical results indicate
that the relaxation times $\tau(\infty)$ are finite for $L\to \infty$
as in the phases $AI/BI$.
As the bulk density in a maximal
current phase is $1/2$ independently of the boundaries,
one could have expected that the relaxation time $\tau(\infty)$ would
be independent of the most efficient reservoir
(i.e. independent of $\beta$ in AII and $\alpha$ in BII).
This is not the case, as boundary layers in the
maximal current domain decay only algebraically,
and lead to a site-dependence of the density.
Besides, strong correlations that cannot be easily characterized
exist in this phase.
This means that 
the maximal current phase cannot be modeled as a flat domain 
with density $1/2$.

But even if we model the maximal current domain as a domain 
of a constant effective density $\rho_{eff}$, which leads 
to the estimated value of $\tau(\infty)$,  we observe large 
deviations between DW predictions and relaxation times 
for finite system sizes - again due to the non trivial
structure of the maximal current domain. The DW predictions would
probably be much improved if one considers site dependent hopping 
rates, but they are difficult to obtain since the structure 
of an isolated maximal current domain is not known.

Anyhow, as in phase AI and BI, the domain wall theory
still offers a simple explanation for
a finite relaxation time of the system in phases AII and BII.

On the disorder line, this picture cannot be applied,
as no domain wall can be identified anymore,
and a qualitatively different dynamical behavior
cannot be excluded. However, though a wall cannot be 
 defined anymore, we could still trace the dynamics 
 of the density fluctuations by introducing a second
 class particle. The second class particle 
 performs a biased motion to one of the boundaries 
 for any density different from $1/2$. Therefore 
 one expects that the density fluctuations are 
 also driven out of the system with a finite velocity, 
 which should lead to finite relaxation times, if 
 $\alpha \neq 1/2$ \cite{Krug91}. 
 
 Contrary, in the periodic system, 
 which has the same simple structure of stationary 
 state, the relaxation times diverge as $L\sim L^{3/2}$
 for arbitrary densities. The divergent relaxation 
 times have been  related to the translational 
 invariance of the periodic system \cite{beijeren} 
 and are, therefore, not expected in case of the open 
 system, except if $\alpha = \beta = 1/2$.
 
 Indeed, our numerical results support this scenario.
Besides, we have shown that the dynamical exponent $z=3/2$ is
also recovered in the whole maximal current phase,
for which  no theoretical prediction exists.

So we could summarize the expected behavior of the relaxation
time as follows~:

(i) In the whole phases AI, AII, BI, and BII, except on the
transition line
$\alpha = \beta$, our results are consistent with a converging
relaxation time.
Its limit value as $L \rightarrow \infty$ is known from
the DW theory in AI and BI.

(ii) On the line $\alpha = \beta$,
and in the whole maximal current phase, the relaxation time diverges
when $L \rightarrow \infty$.
The associated dynamical exponent $z$ is $2$ on the line $\alpha = \beta$
and $3/2$ in the maximal current phase and at the point
$\alpha = \beta = 1/2$.

Further improvement could be obtained from the use
of Finite Size algorithm for DMRG (FSM), as described by
Carlon et al \cite{carlon1},
in order to gain some precision on our calculations - and
thus to reach larger system sizes - in the whole
phase diagram, and especially near critical lines. 

It would also be interesting to apply the DMRG method to other
models, which have a non-trivial but not necessarily site 
dependent domain structure, in order to check whether the DW 
theory still describes correctly the relaxation behavior of the system.

\vskip 1cm

{\bf Acknowledgments}: We acknowledge fruitful discussions with
B. Derrida, J.~Krug and  E.~Carlon, and precious remarks from
G.~Sch\"utz. 
L.S. acknowledges kind hospitality at the L.P.S. and financial
support by C.N.R.S.

\bibliographystyle{unsrt}

\begin{thebibliography}{99}


\bibitem{MacD69}
J.T.~MacDonald and J.H.~Gibbs, Biopolymers {\bf 7}, 707 (1969)

\bibitem{review} D. Chowdhury, L. Santen, and A. Schadschneider,
Phys. Rep. {\bf 329}, 199 (2000)

\bibitem{privman} V.~Privman (ed.), {\em Nonequilibrium Statistical Dynamics 
in One Dimension}, (Cambridge University Press, Cambridge, 1997)

\bibitem{marro} J. Marro, R. Dickman, {\em Nonequilibrium Phase Transitions
in Lattice Models}, (Cambridge University Press, Cambridge, 1999)

\bibitem{gunter} G.M.~Sch\"utz, in: 
{\em Phase Transitions and Critical Phenomena}, Vol. 19, eds. C. Domb
and J.L. Lebowitz (Academic Press, New York,  2000)

\bibitem{spohn} H. Spohn, {\em Large Scale Dynamics of Interacting
Particles}, (Springer, Berlin, 1991) 
 
\bibitem{Krug91}
J.~Krug, Phys. Rev. Lett.~{\bf 67}, 1882 (1991)

\bibitem{Popkov99} V. Popkov and G.M. Sch\"utz, 
Europhys.\ Lett. {\bf 48}, 257 (1999)

\bibitem{Popkov00} V.~Popkov, L.~Santen, A.~Schadschneider, and
G.M.~Sch\"utz, J. Phys. A {\bf 34}, L45 (2001) 

\bibitem{Derr92} B.~Derrida , E.~Domany, and D. Mukamel, J. Stat. Phys.
{\bf 69}, 667 (1992) 


\bibitem{DEHP} B.~Derrida, M.R.~Evans, V.~Hakim, and V.~Pasquier, 
J. Phys. A~{\bf 26}, 1493 (1993)

\bibitem{Gunter93}
 G.M.~Sch\"utz and E.~Domany, J.~Stat.~Phys.~{\bf 72}, 277 (1993)

\bibitem{Derrida} B. Derrida, J.L. Lebowitz and E.R. Speer,
Phys. Rev. Lett. {\bf 87}, 150601 (2001);
B. Derrida, J.L. Lebowitz and E.R. Speer,
cond-mat/0205353.

\bibitem{Gwa} L.-H. Gwa and H. Spohn, Phys. Rev. A {\bf 46}, 844 (1992)

\bibitem{Kim} D.~Kim, Phys. Rev. E, {\bf 52} 3512 (1995)


\bibitem{Kolo98}
A.B.~Kolomeisky, G.M.~Sch\"utz, E.B.~ Kolomeisky, and  J.P.~Straley,
J. Phys.~A {\bf 31}, 6911 (1998)

\bibitem{Belitzky} V. Belitzky and G.M. Sch\"utz, to be published

\bibitem{Bilstein} U. Bilstein and B. Wehefritz, J. Phys. A {\bf 30}, 4925 (1997)


\bibitem{Dud} M. Dudzinski and G.M.~Sch\"utz, J.~Phys. A {\bf 33}, 8351 (2000)



\bibitem{white} S.R. White, Phys. Rev. Lett. {\bf 69}, 2863 (1992)


\bibitem{raj} N. Rajewsky, L. Santen, A. Schadschneider,
          and M. Schreckenberg, J. Stat. Phys. {\bf 92}, 151 (1998)

\bibitem{whitereview} S.R. White,
Phys. Rep. {\bf 301}, 187 (1998)

\bibitem{carlon1} E. Carlon, M. Henkel, and U. Schollw\"{o}ck,
Eur. Phys. J. B {\bf 12}, 99 (1999)

\bibitem{carlon2} E. Carlon, M. Henkel, and U. Schollw\"{o}ck,
Phys. Rev. E {\bf 63}, 036101-1 (2001)

\bibitem{golub} G.H. Golub and C.F. van Loan, {\em Matrix Computations},
3rd edition (Baltimore, 1996)

\bibitem{santen_appert} L. Santen, and C. Appert, 
J. Stat. Phys. {\bf 106}, 187 (2002)

\bibitem{beijeren} H.~v.~Beijeren, 
J. Stat. Phys. {\bf 63}, 47 (1991)
\end{thebibliography}

\end{document}